\documentclass[aps,prl,reprint,superscriptaddress]{revtex4-2}
\usepackage{amsmath}
\usepackage{amssymb}
\usepackage[dvipdfmx]{graphicx}
\usepackage{comment}
\usepackage{color}

\begin{document}


\title{Generation and Annihilation of Topologically-Protected Bound States in the Continuum and Circularly-Polarized States by Symmetry Breaking}


\author{Taiki Yoda}
\affiliation{Department of Physics, Tokyo Institute of Technology, 2-12-1 Ookayama, Meguro-ku, Tokyo 152-8550, Japan}
\affiliation{NTT Basic Research Laboratories, NTT Corporation, 3-1 Morinosato-Wakamiya, Atsugi-shi, Kanagawa 243-0198, Japan}

\author{Masaya Notomi}
\affiliation{Department of Physics, Tokyo Institute of Technology, 2-12-1 Ookayama, Meguro-ku, Tokyo 152-8550, Japan}
\affiliation{NTT Basic Research Laboratories, NTT Corporation, 3-1 Morinosato-Wakamiya, Atsugi-shi, Kanagawa 243-0198, Japan}
\affiliation{Nanophotonics Center, NTT Corporation, 3-1, Morinosato-Wakamiya, Atsugi-shi, Kanagawa 243-0198, Japan}


\date{\today}

\begin{abstract}
We demonstrate by breaking the $C_{6}$ symmetry for higher-order at-$\Gamma$ bound states in the continuum (BICs) with topological charge $-2$ in photonic crystals, (i) deterministic generation of off-$\Gamma$ BICs from the at-$\Gamma$ BIC, and
(ii) a variety of pair-creation and annihilation processes of circularly-polarized states with opposite topological charge and same handedness.
%
These processes are well explained by the conservation of handedness-wise topological charges, showing topologically robustness of these phenomena. 
The results indicate a new control of topological aspect of far-field polarization vectors.
\end{abstract}


\maketitle


\textit{Introduction}.---
The concept of topology has been widely applied to photonic systems~\cite{lu2014topological, lu2016topological, RevModPhys.91.015006}.
Various intriguing phenomena associated with nontrivial 
Bloch wavefunctions are extensively studied.
Recent studies suggest that far-field polarization vector fields radiated from photonic crystals (PhCs), also determined from 
Bloch wavefunctions, possess intriguing topological nature, including bound states in the continuum (BICs)~\cite{hsu2013observation, PhysRevLett.113.257401, hsu2016bound, doeleman2018experimental, PhysRevLett.120.186103, PhysRevLett.122.153907, Jin2019, wang2019generating},
and peculiar reciprocal vortex states~\cite{PhysRevLett.120.186103, zhou2018observation, PhysRevB.99.180101, PhysRevLett.123.116104, chen2019line, yin2019observation}. BICs are resonances with infinite quality factors although their frequencies are inside continuous spectrum of radiation modes.
%
At the $\Gamma$ point, BICs result from the symmetry mismatch between PhC modes and radiative plane waves; this type of BIC is called at-$\Gamma$ BIC or symmetry-protected BIC.
Because of the symmetry protection, the occurrence of at-$\Gamma$ BICs does not depend on structural parameters.

Interestingly, it was found that BICs can also exist at finite wavevectors apart from the symmetry point in a certain type of PhCs, called off-$\Gamma$ BICs or topologically-protected BICs. 
The off-$\Gamma$ BICs have attracted much attention to applications such as on-chip beam steering~\cite{kurosaka2010chip, ha2018directional} and generation of directive vector beams~\cite{PhysRevB.65.195306, iwahashi2011higher, kitamura2012focusing}.
Moreover, off-$\Gamma$ BICs can propagate without radiation loss although their frequency is above the light line~\cite{hu2017propagating, bulgakov2018propagating, gao2017bound}.
In spite of these interesting properties, a systematic way to generate off-$\Gamma$ BICs is still lacking.
The previous off-$\Gamma$ BICs are formed by accidental cancellation of outgoing waves, and thus they are sometimes called \textit{accidental} BICs. With a stark contrast to at-$\Gamma$ BICs, we must tune structural parameters carefully to generate off-$\Gamma$ BICs. As another intriguing point, BICs appear as polarization vortex centers in the reciprocal space~\cite{hsu2013observation, PhysRevLett.113.257401, hsu2016bound, doeleman2018experimental, PhysRevLett.122.153907, PhysRevLett.120.186103, Jin2019, wang2019generating} and carry a topological charge $\nu$ representing the polarization singularity at BICs~\cite{PhysRevLett.113.257401}. Topological protection arises from the conservation of these charges. Theoretically, $\nu$ can be any integer (or half integers for circularly-polarized states (CPSs) which represent another polarization singularity). So far, however, most of previous studies dealt with BICs with $\nu = +1$. 

In this study, we focus at-$\Gamma$ BICs with a higher topological charge, and demonstrate that a variety of interesting phenomena arising from breaking the crystal symmetry of PhCs having a higher-order charge.
First, we demonstrate that this leads to a systematic way to generate off-$\Gamma$ BICs. We show that an at-$\Gamma$ BIC with $\nu = -2$ in triangular-lattice PhCs splits into two off-$\Gamma$ BICs by breaking $C_{6}$ symmetry but preserving $C_{2}$ symmetry. 
These off-$\Gamma$ BICs can be \textit{deterministically} generated by an infinitesimal perturbation without fine-tuning of parameters.
Second, we demonstrate a wide variety of generation and annihilation of CPSs.
We show that a perturbation breaking the $C_{2}$ symmetry but preserving $C_{3}$ symmetry generates six CPSs, and a variety of pair-creation and annihilation of CPSs having \textit{opposite charge} and \textit{same handedness} can occur by symmetry control starting from the at-$\Gamma$ BIC with $\nu = -2$. Importantly, we show that all these phenomena are governed by the conservation of topological charges.
Although previous studies have already pointed out the importance of the charge conservation in far-field topological photonics~\cite{PhysRevLett.113.257401} with an analogy to singular optics for beam propagation~\cite{nye1983lines, dennis2002polarization, bliokh2019geometric}, it is still not clear how far this law can be hold for PhCs.
Actual examples where the charge conservation plays a key role are limited to the case where $\nu $ of BIC is 1 ~\cite{hsu2013observation, PhysRevLett.113.257401, hsu2016bound, doeleman2018experimental,PhysRevLett.122.153907, PhysRevLett.120.186103, Jin2019, wang2019generating, PhysRevLett.120.186103, zhou2018observation, PhysRevB.99.180101, PhysRevLett.123.116104, chen2019line, yin2019observation}, and the applicability of the conservation has not been fully explored.
In this study, we introduce handedness-wise charges ($\nu_\pm$) and clearly proves that they can fully explain a variety of generation and annihilation of BICs and CPSs.     

First of all, let us begin with describing two-dimensional (2D) PhC slabs with a finite thickness~\cite{Joannopoulos:08:Book}.
Above the light line and below the diffraction limit, a non-degenerate eigenmode with an in-plane wavevector ${\bf k_{||}} = (k_{x}, k_{y})$ generally couples to a propagating plane wave with the same ${\bf k}_{||}$ and a polarization vector  ${\bf d}({\bf k}_{||})$ in the $sp$ plane ~\cite{hsu2013observation, PhysRevLett.113.257401}.
To discuss the topology in ${\bf k}_{||}$ space, we introduce a 2D polarization vector projected onto the $xy$ plane,
${\bf d}'({\bf k}_{||}) = d'_{x}({\bf k}_{||}) \hat{x} + d'_{y}({\bf k}_{||}) \hat{y}$
(see Sec.~S1 in the Supplemental Material~\footnote{see Supplemental Material at http:// for (1) the detail of the projected polarization vector, (2) the analysis with the temporal coupled mode theory, (3) the analogy between polarization singularities and band degeneracies, (4) the symmetry consideration of the at-$\Gamma$ BIC, (5) the mobility of the deterministic off-$\Gamma$ BIC, (6) the comparison with the accidental off-$\Gamma$ BIC by fine-tuning, which includes Refs.~\cite{PhysRevB.66.045102, gao2016formation, PhysRevLett.113.257401, fan2003temporal, zhou2016perfect, hsu2017polarization, PhysRevLett.119.167401, PhysRevB.65.235112, liu2016pseudospin, gorlach2018far, parappurath2018direct, guo2019meron, dennis2002polarization, nye1983lines, berry1984quantal, RevModPhys.82.1959, lu2014topological, iwahashi2011higher, hsu2013observation, kurosaka2010chip}}).
The topological charge of the polarization singularities for ${\bf d}'({\bf k}_{||})$ is defined by~\cite{PhysRevLett.113.257401, hsu2016bound, doeleman2018experimental,PhysRevLett.122.153907, zhou2018observation, PhysRevLett.120.186103, PhysRevB.99.180101,Jin2019,wang2019generating, PhysRevLett.123.116104, yin2019observation, chen2019line, dennis2002polarization, bliokh2019geometric, nye1983lines}
\begin{align}
\nu &= \frac{1}{2\pi} \oint_{C} d{\bf k}_{||} \cdot \nabla_{{\bf k}_{||}} \phi({\bf k_{||}}),
\label{topological_charge}
\end{align}
where $C$ is a closed loop in the ${\bf k}_{||}$ space, $\phi({\bf k}_{||}) = \frac{1}{2} \text{arg} [ S_{1}({\bf k}_{||}) + iS_{2}({\bf k}_{||}) ]$ is the angle between its long axis of the polarization ellipse and the $x$ axis, and $S_{i}({\bf k}_{||})$ is the Stokes parameter of ${\bf d}'({\bf k}_{||})$.
$\nu$ describes how many times the polarization ellipse of the polarization vector winds along a loop $C$.
The charge $\nu$ is an integer if a loop $C$ encloses a BIC where $S_{1} = S_{2} = S_{3} = 0$~\cite{PhysRevLett.113.257401, hsu2016bound, doeleman2018experimental, PhysRevLett.120.186103,PhysRevLett.122.153907, Jin2019, wang2019generating}.
The charge $\nu$ is a half-integer if a loop $C$ encloses a CPS where $S_{1} = S_{2} = 0$ and $S_{3} = \pm 1$~\cite{PhysRevLett.123.116104, chen2019line, yin2019observation}.
Equation~(\ref{topological_charge}) is similar to Berry phases in 2D $PT$ symmetric systems (see Sec.~S3 in the Supplemental Material~\cite{Note1}).

\textit{Generation of off-$\Gamma$ BIC from at-$\Gamma$ BIC}.---
The possible charge of an at-$\Gamma$ BIC is determined by the eigenvalues of the rotational symmetry of the system~\cite{PhysRevLett.113.257401}.
In PhCs with $C_{2}$ symmetry and without $C_{4}$ and $C_{6}$ symmetry, at-$\Gamma$ BICs with $|\nu | \geq 2$ are not allowed, but they become allowed when PhCs posses $C_{4}$ or $C_{6}$ symmetry.
Therefore, an at-$\Gamma$ BIC with $|\nu | \geq 2$ splits into multiple off-$\Gamma$ BICs through the charge conservation when a perturbation breaking $C_{4}$ or $C_{6}$ symmetry but preserving $C_{2}$ symmetry is introduced (see Sec.~S4 in the Supplemental Material~\cite{Note1}).
Since this splitting relies on the charge conservation and crystalline symmetry, the generation of off-$\Gamma$ BICs can be controlled by systematic deformation of the lattice symmetry in contrast to the generation of off-$\Gamma$ BICs by an accidental fine-tuning of structural parameters~\cite{PhysRevLett.113.257401}.
It is worth noting that although the destruction of at-$\Gamma$ BICs by reducing symmetry has been reported~\cite{cui2012dynamic, overvig2019selection}, the transition from an at-$\Gamma$ BIC to off-$\Gamma$ BICs has not been pointed out in previous works.

\begin{figure}
\includegraphics[width=8.6cm]{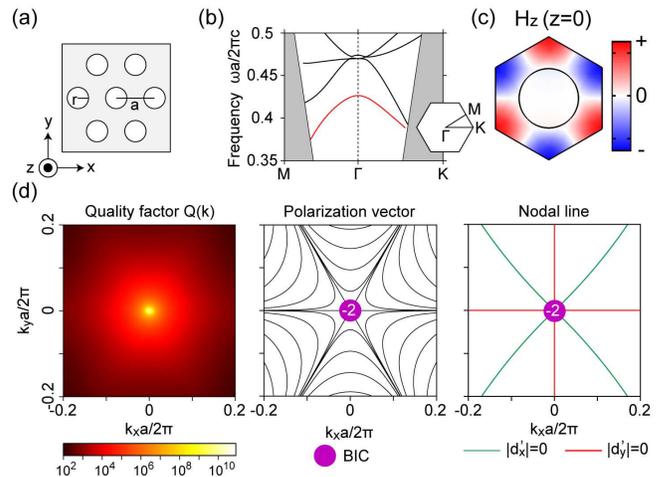}%
\caption{
(a) Schematics of a PhC slab with a triangular lattice of circular air-holes.
(b) Calculated TE-like band structure.
The red line is the lowest TE-like band.
The gray region indicates a region below the light line.
The inset is the first Brillouin zone.
(c) Mode profile in the unit cell for the lowest TE-like band at the $\Gamma$ point.
%
(d) Calculated quality factor (left), polarization vector (middle), and nodal lines of $d'_{x}$ and $d'_{y}$ (right).
The projected polarization vector is represented by the line field tangent to the long axis of the polarization ellipse~\cite{PhysRevLett.120.186103, Jin2019, wang2019generating, zhou2018observation, PhysRevB.99.180101, PhysRevLett.123.116104, yin2019observation, chen2019line, nye1983lines, dennis2002polarization}.
}
\label{fig_triangular_lattice}
\end{figure}

To verify the splitting of the at-$\Gamma$ BIC, we focus on eigenmodes whose $C_{6}$ eigenvalue is $-1$ at the $\Gamma$ point because they always result in at-$\Gamma$ BICs with $|\nu | \geq 2$~\cite{PhysRevLett.113.257401}.
We numerically investigate a triangular-lattice PhC slab of circular air-holes with lattice constant $a$, hole radius $r=0.27844a$, slab thickness $h=0.38a$, and refractive index $n=3.48$ [Fig.~\ref{fig_triangular_lattice}(a)] by the finite-element method.
The mode profile of the lowest TE-like band [Fig.~\ref{fig_triangular_lattice}(b)] at the $\Gamma$ point is shown in Fig.~\ref{fig_triangular_lattice}(c).
This mode belongs to the representation $B_{1}$ of the $C_{6v}$ point group, and thus becomes an at-$\Gamma$ BIC with charge $-2+6n$~\cite{PhysRevLett.113.257401}.
The calculated quality factor ($Q$) and polarization vector of the lowest TE-like band is plotted in Figs.~\ref{fig_triangular_lattice}(d).
We find that $Q$ diverges at the $\Gamma$ point and that the polarization vector winds twice around it.
We also plot nodal lines of $d'_{x}$ and $d'_{y}$ in Fig.~\ref{fig_triangular_lattice}(d),
and find that both nodal lines are doubly degenerate at the $\Gamma$ point.
These results demonstrate that a BIC with charge -2 exists at the $\Gamma$ point~\cite{PhysRevB.65.195306, PhysRevLett.113.257401, PhysRevLett.122.153907}.

\begin{figure}
\includegraphics[width=8.6cm]{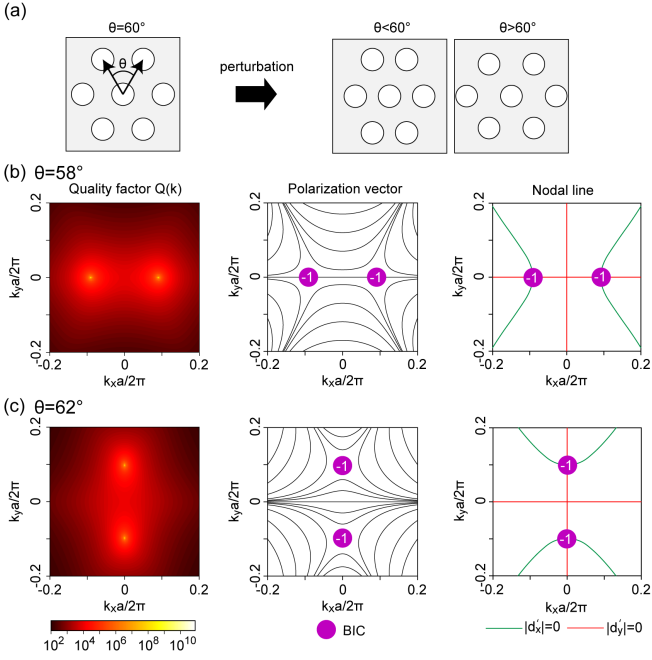}%
\caption{
(a) Definition of the angle $\theta$ and schematics of perturbed PhCs.
(b)(c) Calculated quality factor (left), polarization vector (middle), and nodal lines of $d'_{x}$ and $d'_{y}$ (right) for (b) $\theta = 58^{\circ}$ and (c) $\theta = 62^{\circ}$.
}
\label{fig_sheared_triangular_lattice}
\end{figure}

Next, we introduce perturbations breaking the $C_{6}$ symmetry but preserving the $C_{2}$ symmetry.
Here we slightly vary the angle $\theta$ between two translational vectors defined in Fig.~\ref{fig_sheared_triangular_lattice}(a)~\cite{PhysRevB.85.035304}.
When $\theta$ is deviated from 60$^\circ$ (corresponding to uniaxial deformation in the triangular lattice), the symmetry of the eigenmode for the lowest TE-like band at the $\Gamma$ point is reduced to the $B_{1}$ representation of the $C_{2v}$ point group.
The perturbed eigenmode at the $\Gamma$ point is no longer a symmetry-protected BIC.
Therefore, we can expect that the perturbation will split the original at-$\Gamma$ BIC with charge -2 into two off-$\Gamma$ BICs with charge -1 through the charge conservation.
We numerically calculate $Q$ for $\theta = 58^\circ$ and $\theta = 62^\circ$ (Fig.~\ref{fig_sheared_triangular_lattice}(b)(c) left).
We can observe two high $Q$ states at finite wavevectors on a mirror-invariant axis ($k_{x} = 0$ or $k_{y} = 0$).
To prove that these states are exact BICs, not quasi-BICs~\cite{PhysRevLett.121.193903, PhysRevApplied.12.014024}, we also plot the polarization vector and nodal lines of $d'_{x}$ and $d'_{y}$ in Fig.~\ref{fig_sheared_triangular_lattice}(b)(c).
The polarization ellipse winds by $-2\pi$ around the high-$Q$ states (Fig.~\ref{fig_sheared_triangular_lattice}(b)(c) middle).
This plot shows that the perturbation splits the degeneracy of the nodal lines of $d'_{x}$ and $d'_{y}$ at the $\Gamma$ point.
In addition, the nodal lines of $d'_{x}$ and $d'_{y}$ intersect each other at the high-$Q$ states (Fig.~\ref{fig_sheared_triangular_lattice}(b)(c) right), where the polarization vector vanishes~\cite{hsu2013observation, PhysRevLett.113.257401, PhysRevLett.119.167401}.
These results unambiguously verify that the at-$\Gamma$ BIC with charge $-2$ splits into the two exact off-$\Gamma$ BICs with charge $-1$ by the uniaxial deformation.
This finding shows that there is always a systematic way to generate off-$\Gamma$ BICs from higher-order at-$\Gamma$ BICs.

Conventional off-$\Gamma$ BICs can exist only at a certain range of the structural parameters.
In Sec.~S5 in the Supplemental Material~\cite{Note1}, we have shown an example that off-$\Gamma$ BICs does not exist when $h < 0.98a$ for triangular-lattice PhCs with $r=0.27844a$, $n=3.48$, and $\theta=60^{\circ}$.
This contrasts with the present deterministic off-$\Gamma$ BICs which always exist at any thickness because the existence of at-$\Gamma$ BIC is guaranteed by the symmetry. 
Furthermore, we have confirmed that it is possible to widely change the wavevector of off-$\Gamma$ BIC by simply varying $\theta$ as shown in Sec.~S6 in the Supplemental Material~\cite{Note1}.
The result shows that one can widely change the emission angle of polarization vortex beams generated from BICs between 0 and 90 degree, which will be promising for vortex-beam steering devices. 

\begin{figure}
\includegraphics[width=8.6cm]{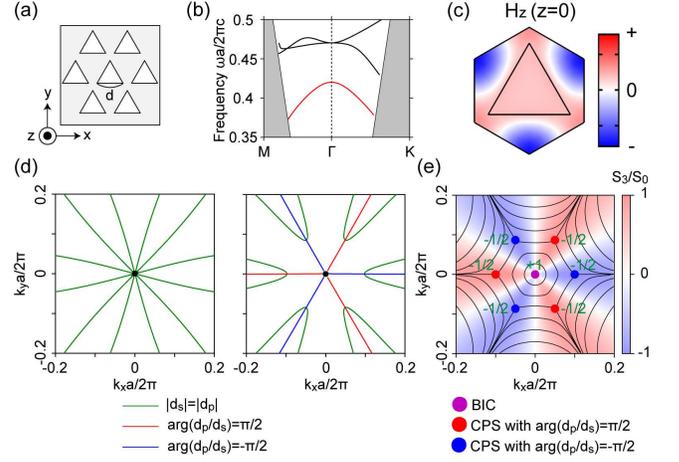}%
\caption{
(a) A triangular-lattice PhC with triangular air-holes.
(b) Calculated TE-like band structure.
(c) Mode profile in the unit cell for the lowest TE-like band at the $\Gamma$ point.
(d) Lines for $|d_{s}| = |d_{p}|$ (green) and $\delta = \pm \pi/2$ (red and blue) for PhCs with the circular (left) and triangular (right) holes.
(e) Calculated polarization vector and normalized Stokes parameter $S_{3}/S_{0} = 2|d_{s}||d_{p}|\sin\delta/(|d_{s}|^{2}+|d_{p}|^{2})$.
The attached numbers denote topological charges $\nu$.
}
\label{fig_triangular_lattice_with_triangular_hole}
\end{figure}

\begin{figure*}
\includegraphics[width=17.2cm]{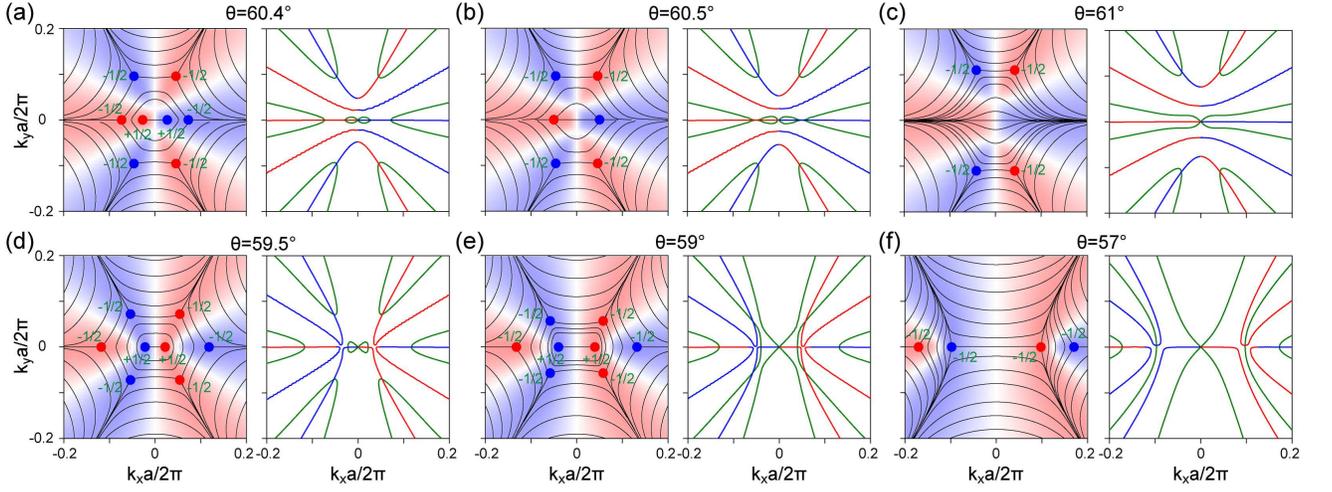}%
\caption{
Evolution of CPSs.
In (a)-(f), left panels show calculated polarization vector and normalized Stokes parameter $S_{3}/S_{0}$, and right panels show lines for $|d_{s}| = |d_{p}|$ (green) and $\delta = \pm \pi/2$ (red and blue).
For (a) $\theta=60.4^{\circ}$ and (d) $\theta = 59.5^{\circ}$, the at-$\Gamma$ BIC ($\nu = +1$) split into two CPSs ($\nu = +1/2$).
For (b) $\theta=60.5^{\circ}$ and (e) $\theta = 59^{\circ}$, two CPSs with opposite charges and same handedness approach each other.
For (c) $\theta = 61^{\circ}$ and (f) $\theta = 57^{\circ}$, two CPSs with opposite charges and same handedness collide, and then they annihilate by the collision.
In (a)-(f), one of five lines for $\delta = \pm 1/2$ is pinned at the $\Gamma$ point due to $\sigma T$ symmetry (see Sec.~S2 in the Supplemental Material~\cite{Note1}).
}
\label{fig_sheared_triangular_lattice_with_triangular_hole}
\end{figure*}
%
%
\textit{Generation and annihilation of CPSs}.---In PhC with $C_{2}$ symmetry, the polarization vector is mostly linear-polarized~\cite{hsu2017polarization, Note1} and it is generally hard to obtain CPSs.
%
%
%
Recently, generation of two CPSs from splitting a BIC with charge $\pm 1$ in PhCs by breaking spatial symmetries was reported~\cite{yin2019observation, PhysRevLett.123.116104, chen2019line} .
Here we demonstrate that starting from higher-order at-$\Gamma$ BIC with charge -2, we are able to observe far more rich phenomena including various pair-creation and annihilation of CPSs.

Before showing the actual examples, we point out that $\nu$ defined by Eq.~(\ref{topological_charge}) cannot fully describe the processes involved with CPSs because it does not contain the Stokes parameter $S_{3}$.
To overcome this difficulty, we express ${\bf d}'({\bf k}_{||})$ in terms of the circular basis~\cite{berry1987adiabatic, dennis2002polarization,bliokh2019geometric},
\begin{equation}
{\bf d}'({\bf k}_{||}) = d_{+}'({\bf k}_{||}) {\bf e}_{+} + d_{-}'({\bf k}_{||}) {\bf e}_{-},
\end{equation}
where ${\bf e}_{\pm} = (\hat{x} \pm i \hat{y})/\sqrt{2}$ and $d'_{\pm}({\bf k}_{||}) = [ d'_{x}({\bf k}_{||}) \mp i d'_{y}({\bf k}_{||}) ]/\sqrt{2}$.
In this basis, zeros of $d'_{\pm}({\bf k}_{||})$ correspond to CPSs.
When both $d'_{+}({\bf k}_{||})$ and $d'_{-}({\bf k}_{||})$ are simultaneously zero, a eigenmode turns into a BIC.
By writing $d'_{\pm}({\bf k}_{||})$ as $d'_{\pm}({\bf k}_{||}) = |d'_{\pm}| e^{i \alpha_{\pm}}$, we introduce an integer~\cite{dennis2002polarization, bliokh2019geometric},
\begin{equation}
\nu_{\pm} = \frac{1}{2\pi} \oint_{C} d {\bf k}_{||} \cdot \nabla_{{\bf k}_{||}} \alpha_{\pm} ({\bf k}_{||}).
\label{Npm}
\end{equation}
$\nu_{\pm}$ is a finite integer when a loop $C$ encloses the zeros of $d'_{\pm}({\bf k}_{||})$.
$\nu$ and $\nu_{\pm}$ are related by
\begin{align}
    \nu = \frac{1}{2}(\nu_{-} - \nu_{+})
    \label{nu_and_nu_pm}
\end{align}
because $\phi({\bf k}_{||})$ is given by $\phi({\bf k}_{||}) = \frac{1}{2} \text{arg} (d'^{*}_{+}d'_{-})$.
It follows from Eq.~(\ref{nu_and_nu_pm}) that $\nu_{\pm}$ is a handedness-wise charge, meaning the winding number weighted by its handedness $S_{3}$.
Assuming the conservation for $\nu$ and $\nu_{\pm}$, a BIC with $\nu = \pm 1$ can be regarded as the superposition of two CPSs with $(\nu_{+}, \nu_{-}) = \pm(-1, +1)$, which indeed correctly explains the observation that a perturbation breaking $C_{2}$ symmetry splits the BIC into two CPSs with \textit{same topological charge} $\nu$ and \textit{opposite handedness} $S_{3}$~\cite{PhysRevLett.123.116104, chen2019line, yin2019observation}.
Furthermore, a CPS with $\nu_{\pm} = +1$ can be removed through collision with another CPS with $\nu_{\pm} = -1$, i.e., \textit{opposite topological charge} $\nu$ and \textit{same handedness} $S_{3}$.
The introduction of $\nu_{\pm}$ consistently explains the splitting from a BIC into two CPSs and predicts the rule for pair-annihilation of two CPSs.

%
%

%
Here we examine actual examples. We break the $C_{2}$ symmetry by deforming air holes into a triangular shape with a side length of $d=0.75a$ and a slab thickness of $h=0.38a$ [Fig.~\ref{fig_triangular_lattice_with_triangular_hole}(a)].
This deformation reduces the crystalline symmetry from $C_{6v}$ to $C_{3v}$, which allows CPSs.
The perturbation does not break the at-$\Gamma$ BIC because $C_{3}$ symmetry still protects an at-$\Gamma$ BIC with charge $1+3n$~\cite{PhysRevLett.113.257401}.
We plot the calculated polarization vector for the lowest TE-like band [Fig~\ref{fig_triangular_lattice_with_triangular_hole}(e)].
The polarization ellipse now winds only once in the opposite direction around the $\Gamma$ point, indicating $\nu$ of the at-$\Gamma$ BIC changes from $-2$ to $+1$.
Since the present PhC is no longer invariant under $C_{2}$ symmetry, the change of $\nu$ may generate six CPSs with $\nu = -1/2$ through the charge conservation.
This prediction is indeed confirmed in Fig.~\ref{fig_triangular_lattice_with_triangular_hole}(e). We find that the polarization ellipse winds by $-\pi$ around six CPSs.
The half-winding indicates $\nu = -1/2$.
It should be noted that among the six CPSs, three CPSs are left-handed with $\delta = \pi/2$ ($\delta \equiv \text{arg}(d_{p}/d_{s})$) and the others are right-handed with $\delta = -\pi/2$, which is explained by the conservation of $\nu_{\pm}$.

For better understanding, we plot lines for $|d_{s}| = |d_{p}|$ (Lines A) and $\delta = \pm \pi/2$ (Lines B).
For the circular-hole PhC (Fig.~\ref{fig_triangular_lattice_with_triangular_hole}(d) left), Lines A emerge between the lines for $d_{s}=0$ and $d_{p}=0$~\cite{PhysRevLett.119.167401}.
Lines B do not exist due to the $C_{2}$ symmetry.
For the triangular-hole PhC (Fig.~\ref{fig_triangular_lattice_with_triangular_hole}(d) right), the degeneracy of Lines A at the $\Gamma$ point is lifted, and Lines B emerge.
The six intersections between Lines A and B correspond to CPSs, which guarantees the topological protection of CPSs in a similar way to that of BICs~\cite{PhysRevLett.113.257401}.

Finally, we vary the angle $\theta$ from $60^{\circ}$ to introduce the uniaxial deformation.
It is expected from the conservation of $\nu_{\pm}$ that perturbations breaking the $C_{3}$ symmetry splits the at-$\Gamma$ BIC into two CPSs with same charge $+1/2$ and opposite handedness [Fig.~\ref{fig_sheared_triangular_lattice_with_triangular_hole}(a)(d)].
For $\theta > 60^{\circ}$ ($\theta < 60^{\circ}$), the CPS with $\delta = -\pi/2$ splits in the direction of $k_{x} > 0$ ($k_{x} < 0$).
As the perturbation is increased, two CPSs with opposite charge and same handedness approaches [Fig.~\ref{fig_sheared_triangular_lattice_with_triangular_hole}(b)(e)], and eventually collide each other, resulting pair-annihilation of CPSs [Fig.~\ref{fig_sheared_triangular_lattice_with_triangular_hole}(c)(f)].
Finally, the two pair of CPSs with charge $-1/2$ and opposite handedness survive.
Note that these two pair of CPSs can be regarded as generated from the off-$\Gamma$ BICs in Fig.~\ref{fig_sheared_triangular_lattice}(b)(c) as a result of breaking $C_{2}$ symmetry.
Let us point out that all through these process [Fig.~\ref{fig_sheared_triangular_lattice_with_triangular_hole}(a)-(f)], the topological charges $\nu_{\pm}$ are consistently conserved, and proved that conventional topological charges $\nu$ are not sufficient for explaining the results.

%
%
%
In conclusion, starting from a single at-$\Gamma$ BIC with charge $-2$, we have demonstrated by breaking its $C_{6}$ symmetry, deterministic generation of off-$\Gamma$ BICs, and various forms of generation and pair-annihilation of CPSs.
All processes are successfully explained by the charge conservation for handedness-wise topological charges $\nu_{\pm}$.
This clearly demonstrates the wide applicability of the charge conservation of $\nu_{\pm}$ for far-field topological photonics by PhCs, and guarantees the stability of their polarization singularities in parameter spaces.
Furthermore, our findings pave the novel way for various polarization singularity control of far-field radiations such as circularly-polarized light emission~\cite{PhysRevLett.106.057402, PhysRevB.89.045316, PhysRevB.92.205309}.
Especially, this novel way for generating off-$\Gamma$ BICs will be promising for on-chip vortex-beam steering~\cite{kurosaka2010chip, ha2018directional} and dynamic tuning of a quality factor~\cite{cui2012dynamic}.

\begin{acknowledgments}
We thank Yuto Moritake and Kenta Takata for helpful discussions.
This work was supported by Japan Society for the Promotion of Science (KAKENHI JP15H05735).
\end{acknowledgments}

\end{document}



\title{Supplemental material for ``Generation and Annihilation of Topologically-Protected Bound States in the Continuum and Circularly-Polarized States by Breaking Symmetry"}

\date{\today}



\maketitle

\renewcommand{\theequation}{S\arabic{equation}}
\renewcommand{\thefigure}{S\arabic{figure}}
\renewcommand{\bibnumfmt}[1]{[S#1]}
\renewcommand{\citenumfont}[1]{S#1}

\section{S1. Detail of polarization vector}
In two-dimensional PhCs with a finite thickness, the periodicity in the $xy$ direction ensures that the electric field of a resonance above the light line can be written as ${\bf E}_{{\bf k}_{||}}(\boldsymbol \rho, z) = e^{-i {\bf k}_{||} \cdot \boldsymbol \rho} {\bf u}_{{\bf k}_{||}}(\boldsymbol \rho, z)$, where ${\bf k}_{||} = (k_{x}, k_{y})$ is the in-plane wavevector, $\boldsymbol \rho = (x, y)$ is the in-plane coordinate, ${\bf u}_{{\bf k}_{||}}(\boldsymbol \rho, z)$ is the periodic function in $\boldsymbol \rho$, and $z$ is the normal coordinate perpendicular to the slabs.
The resonance above the light line is a solution of the Maxwell equation with complex frequency $\omega({\bf k}_{||}) = \omega_{0} + i\gamma$ and an outgoing boundary condition~\cite{PhysRevB.66.045102, gao2016formation}.
Above the light line and below the diffraction limit, a non-degenerate resonance at ${\bf k}_{||}$ couples to a radiative plane wave with the in-plane wavevector ${\bf k}_{||}$ and a polarization vector ${\bf d}^{u, d}({\bf k}_{||}) = d_{x}^{u, d}({\bf k}_{||}) \hat{x} + d_{y}^{u, d}({\bf k}_{||}) \hat{y} + d_{z}^{u, d}({\bf k}_{||}) \hat{z}$.
The superscripts $u$ and $d$ denote above and below the slabs respectively.
The polarization vector is given by the zero-order Fourier components of Bloch functions: $d_{x}^{u,d}({\bf k}_{||}) = \hat{x} \cdot \langle {\bf u}_{{\bf k}_{||}} \rangle$,  $d_{y}^{u,d}({\bf k}_{||}) = \hat{y} \cdot \langle {\bf u}_{{\bf k}_{||}} \rangle$, and  $d_{z}^{u,d}({\bf k}_{||}) = \hat{z} \cdot \langle {\bf u}_{{\bf k}_{||}} \rangle$, where the braket denotes spatial averaging over a unit cell above and below the slabs~\cite{PhysRevLett.113.257401, gao2016formation}.
Because the radiative plane wave is transverse, the polarization vector is decomposed into two orthogonal components: ${\bf d}^{u,d}({\bf k}_{||}) = d^{u, d}_{s}({\bf k}_{||})\hat{s} + d^{u, d}_{p}({\bf k}_{||})\hat{p}$, where $\hat{s} = \hat{z} \times \hat{k} / |\hat{z} \times \hat{k}|$, $\hat{z}$ is the upward unit vector perpendicular to the slabs, $\hat{k}$ is the outgoing unit vector parallel to the wavevector, $\hat{p} = \hat{k} \times \hat{s}$, $d^{u, d}_{s}({\bf k}_{||}) = \hat{s} \cdot \langle {\bf u}_{{\bf k}_{||}} \rangle$ and $d^{u, d}_{p}({\bf k}_{||}) = \hat{p} \cdot \langle {\bf u}_{{\bf k}_{||}} \rangle$.
As discussed later, the superscript $u$ and $d$ can be omitted when systems have a mirror symmetry with respect to the $xy$ plane ($\sigma_{z}$ symmetry).
We assume $\sigma_{z}$ symmetry throughout the main text.

Generally the polarization vector is a three-dimensional vector on the $sp$ plane.
To discuss the topology of the polarization vector, we should define a two-dimensional polarization vector projected onto the $xy$ plane.
In principle, there are several ways to project the polarization vector ${\bf d}({\bf k}_{||})$ onto the $xy$ plane.
In Ref.~\cite{PhysRevLett.113.257401}, a projected polarization vector
\begin{align}
{\bf c}({\bf k}_{||}) &= d_{x}({\bf k}_{||}) \hat{x} + d_{y}({\bf k}_{||}) \hat{y}
\label{def_polarization_vector}
\end{align}
was introduced.
However, the projection (\ref{def_polarization_vector}) does not maintain the polarization state.
To discuss the polarization vector on the $sp$ plane, we need to introduce projections which maintain the polarization state of ${\bf d}({\bf k}_{||})$.
In this letter, we define a two-dimensional polarization vector projected onto the $xy$ plane,
\begin{align}
{\bf d}'({\bf k}_{||}) &= d'_{x}({\bf k}_{||}) \hat{x} + d'_{y}({\bf k}_{||}) \hat{y}
\\
&=
\begin{cases}
	d_{s}({\bf k}_{||}) \hat{s}' + d_{p}({\bf k}_{||}) \hat{p}' & {\bf k}_{||} \neq {\bf 0}, \\
	d_{x}({\bf k}_{||}) \hat{x} + d_{y}({\bf k}_{||}) \hat{y}  & {\bf k}_{||} = {\bf 0},
\end{cases}
\end{align}
where $\hat{s}' = \hat{z} \times \hat{k}_{||}$, $\hat{p}' = \hat{z} \times \hat{s}'$, and $\hat{k}_{||} = (k_{x}, k_{y})/\sqrt{k_{x}^{2} + k_{y}^{2}}$.
This projection clearly maintain the shape of the polarization ellipse.
Note that we should use the polarization vector ${\bf c}({\bf k}_{||})$ if we are interested in the $xy$ components of the polarization vector.
The theory developed in the main text can be applied to the polarization vector ${\bf c}({\bf k}_{||})$.

\section{S2. Temporal coupled mode theory}
\begin{figure}
\includegraphics[width=8.6cm]{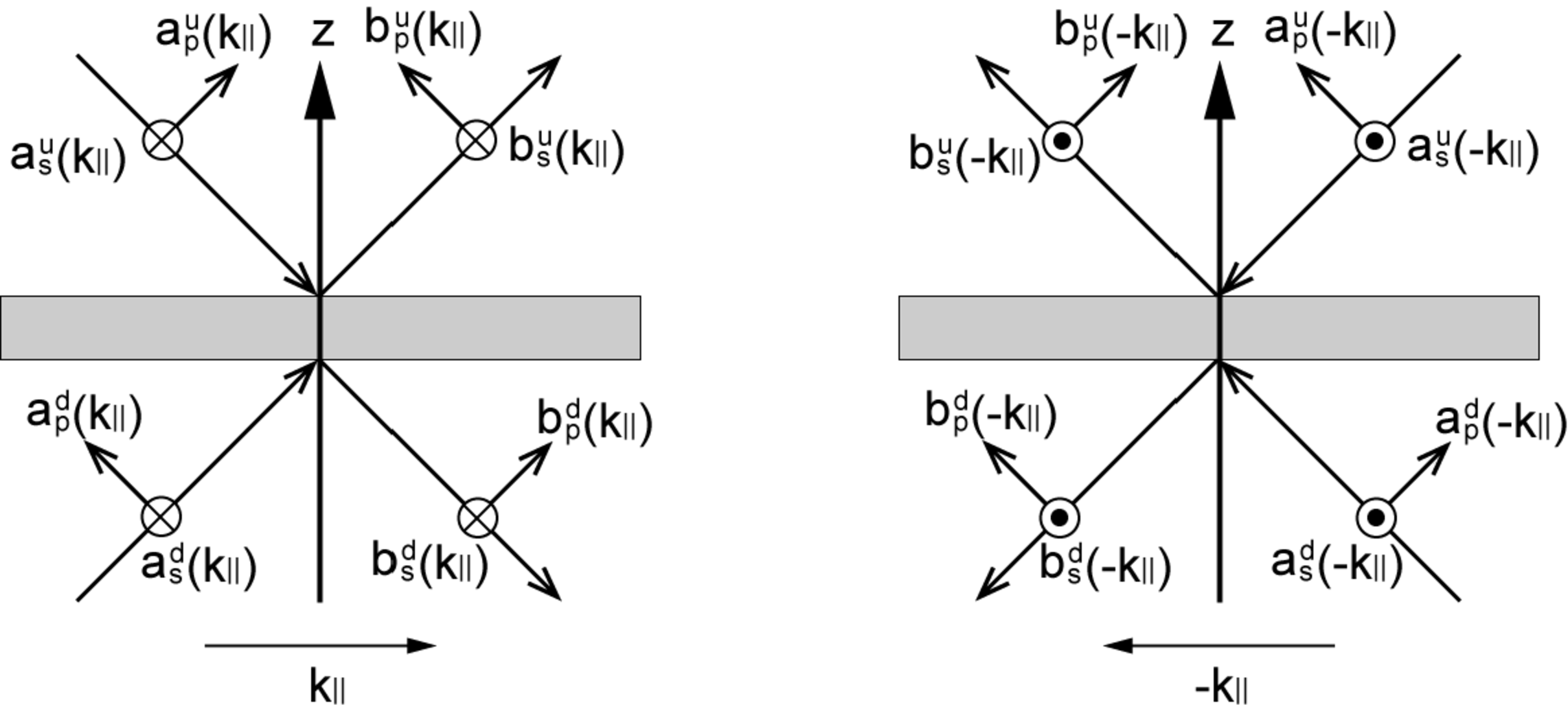}%
\caption{
Definition of input and output waves with $\hat{s} = \hat{z} \times \hat{k} / |\hat{z} \times \hat{k}|$ and $\hat{p} = \hat{k} \times \hat{s}$.
The direction of $\hat{s}$ is opposite for the process at ${\bf k}_{||}$ and $-{\bf k}_{||}$, while the direction of $\hat{p}$ is same for the process at ${\bf k}_{||}$ and $-{\bf k}_{||}$.
}
\label{fig_tcmt}
\end{figure}
In this section, we briefly review a temporal coupled-mode theory (TCMT)~\cite{fan2003temporal,zhou2016perfect,hsu2017polarization,PhysRevLett.119.167401}.
Because the sign convention we use here is a little different from previous works, we give a detailed derivation of the TCMT in this section.
Consider incident waves with an in-plane wavevector ${\bf k}_{||}$ and outgoing waves with the in-plane wavevector ${\bf k}_{||}$ (Fig.~\ref{fig_tcmt}).
The input and output waves are associated with a scattering matrix $S({\bf k}_{||})$,
\begin{equation}
b({\bf k}_{||}) = S({\bf k}_{||}) a({\bf k}_{||}),
\label{scattering_matrix}
\end{equation}
where $a({\bf k}_{||}) = (a^{u}_{s}({\bf k}_{||}), a^{d}_{s}({\bf k}_{||}), a^{u}_{p}({\bf k}_{||}), a^{d}_{p}({\bf k}_{||}))^{T}$ is the column vector containing the complex amplitude of the incident wave, $b({\bf k}_{||}) = (b^{u}_{s}({\bf k}_{||}), b^{d}_{s}({\bf k}_{||}), b^{u}_{p}({\bf k}_{||}), b^{d}_{p}({\bf k}_{||}))^{T}$ is the column vector containing the complex amplitude of the outgoing wave as defined in Fig.~\ref{fig_tcmt}.
The superscript $u$ $(d)$ denotes above (below) the slab, and the subscript denotes the polarization.
The scattering matrix is restricted by energy conservation and time-reversal symmetry.
When a PhC is lossless, the energy conservation guarantees that the outgoing power equals to the incident power:
\begin{equation}
a^{\dagger}({\bf k}_{||})a({\bf k}_{||}) = b^{\dagger}({\bf k}_{||})b({\bf k}_{||}).
\label{energy_conservation}
\end{equation}
By inserting Eq.~(\ref{scattering_matrix}) into Eq.~(\ref{energy_conservation}), we can prove the unitarity of S:
\begin{equation}
S^{-1}({\bf k}_{||}) = S^{\dagger}({\bf k_{||}}).
\label{unitarity}
\end{equation}
Next, we consider the time-reversed process of Eq.~(\ref{scattering_matrix}).
Let $a'(-{\bf k}_{||})$ be the complex amplitude of the time-reversed wave of $b({\bf k}_{||})$, and let $b'(-{\bf k}_{||})$ be the complex amplitude of the time-reversed wave of $a({\bf k}_{||})$.
The time-reversed operation is represented by complex conjugation up to an arbitrary phase factor.
Moreover, the time-reversal operation does not flip the direction of the electric field, while the time-reversal operation flips the direction of $\hat{s}$ because $\hat{s}$ depends on the direction of ${\bf k}_{||}$.
As a result, the time-reversal operation flips the sign of the $s$ component in our coordinate system (Fig.~\ref{fig_tcmt})~\cite{PhysRevLett.119.167401}.
Therefore, the time-reversed wave are expressed as $a'(-{\bf k}_{||}) = -e^{i\theta}(\tau_{z} \otimes I) b^{*}({\bf k}_{||})$ and $b'(-{\bf k}_{||}) = -e^{i\theta}(\tau_{z} \otimes I) a^{*}({\bf k}_{||})$, where $\tau_{i}$ is the Pauli matrix acting on the polarization space, $I$ is the identity matrix, and $e^{i\theta}$ is an arbitrary phase factor.
Because the time-reversed process is identified with a process with in-plane wavevector $- {\bf k}_{||}$, they need to satisfy
\begin{equation}
b'(-{\bf k}_{||}) = S(-{\bf k}_{||})a'(-{\bf k}_{||}).
\label{time-reversed}
\end{equation}
By inserting $a'(-{\bf k}_{||}) = -e^{i\theta}(\tau_{z} \otimes I) b^{*}({\bf k}_{||})$ and $b'(-{\bf k}_{||}) = -e^{i\theta}(\tau_{z} \otimes I) a^{*}({\bf k}_{||})$ into Eq.~(\ref{time-reversed}), we obtain a constraint imposed by the time-reversal symmetry
\begin{equation}
S({\bf k}_{||}) = (\tau_{z} \otimes I) S^{T}(-{\bf k}_{||}) (\tau_{z} \otimes I).
\label{time-reversal_symmetry}
\end{equation}
%

The dynamics of a non-degenerate resonance in a PhC is described by the TCMT~\cite{fan2003temporal},
\begin{align}
\frac{dA({\bf k}_{||})}{dt}  &= (i\omega_{0} - \gamma) A({\bf k}_{||}) + K^{T}({\bf k}_{||}) a({\bf k}_{||}),
\label{tcmt_eq1}
\\
b({\bf k}_{||}) &= C({\bf k}_{||}) a({\bf k}_{||}) + D({\bf k}_{||}) A({\bf k}_{||}),
\label{tcmt_eq2}
\end{align}
where $A({\bf k}_{||})$ is the complex amplitude of the resonance at ${\bf k}_{||}$,
$K({\bf k}_{||}) = (k^{u}_{s}, k^{d}_{s}, k^{u}_{p}, k^{d}_{p})^{T}$ is the coupling constants from the incident wave to the resonance, and
$D({\bf k}_{||}) = (d^{u}_{s}, d^{d}_{s}, d^{u}_{p}, d^{d}_{p})^{T}$ is the coupling constants from the resonance to the outgoing wave.
The matrix $C$ is a scattering matrix describing the direct scattering process, and approximates the scattering matrix of a uniform dielectric slab~\cite{PhysRevB.65.235112, hsu2017polarization}:
\begin{equation}
C =
	\begin{pmatrix}
		C_{s} && 0 \\
		0 && C_{p}
	\end{pmatrix},
C_{s, p} =
	\begin{pmatrix}
		r_{s, p} && t_{s, p} \\
		t_{s, p} && r_{s, p}
	\end{pmatrix}.
\end{equation}
We assume that the directi scattering process does not convert the polarization and that the direct scattering matrix $C$ satisfies Eq.~(\ref{unitarity}) and Eq.~(\ref{time-reversal_symmetry}).
When the resonance is excited by external incident waves with in-plane wavevector ${\bf k}_{||}$ and frequency $\omega$, the scattering matrix is written by
\begin{equation}
S({\bf k}_{||}) = C + \frac{D({\bf k}_{||}) K^{T}({\bf k}_{||})}{i(\omega - \omega_{0})+\gamma}.
\label{scattering_matrix_TCMT1}
\end{equation}

The constant $K, D$ and $C$ are related each other via energy conservation and time-reversal symmetry.
Inserting Eq.~(\ref{scattering_matrix_TCMT1}) into Eq.~(\ref{time-reversal_symmetry}) yields
\begin{align}
\omega_{0}({\bf k}_{||}) &= \omega_{0}(-{\bf k}_{||}),
\\
\gamma({\bf k}_{||}) &= \gamma(-{\bf k}_{||}),
\\
D({\bf k}_{||})K^{T}({\bf k}_{||}) &= (\tau_{z} \otimes I) K(-{\bf k}_{||})D^{T}(-{\bf k}_{||}) (\tau_{z} \otimes I)
\label{unitarity2}.
\end{align}
Next, consider a process where the incident wave is absent and the resonance decays exponentially into the outgoing wave.
In this process, the energy conservation guarantees
\begin{equation}
\frac{d}{dt}|A({\bf k}_{||})|^{2} = - b^{\dagger}({\bf k}_{||})b({\bf k}_{||}).
\label{energy_conservation2}
\end{equation}
By combining Eqs.~(\ref{tcmt_eq1}), (\ref{tcmt_eq2}) and (\ref{energy_conservation2}) under the condition of $a = 0$, we obtain
\begin{equation}
D^{\dagger}({\bf k}_{||})D({\bf k}_{||}) = 2\gamma({\bf k}_{||}).
\label{energy_conservation3}
\end{equation}
Next, we consider the time-reversed process of the decay process where the resonance exponentially grow without the outgoing wave.
In this process, the incident wave $a'(-{\bf k}) = - e^{i\theta}(\tau_{z} \otimes I) D^{*}({\bf k}_{||})A^{*}({\bf k}_{||})$ excites the resonance $A'({-\bf k}_{||}) (=e^{i\theta}A^{*}({\bf k}_{||}))$ with complex frequency $\omega_{0} - i\gamma$.
By inserting these conditions into Eq.~(\ref{tcmt_eq1}), we obtain
\begin{equation}
2\gamma({\bf k}_{||}) = - K^{T}(-{\bf k}_{||}) (\tau_{z} \otimes I) D^{*}({\bf k}_{||}).
\label{K-D}
\end{equation}
Combining Eqs.~(\ref{unitarity2}), (\ref{energy_conservation3}), and (\ref{K-D}) produces
\begin{equation}
K({\bf k}_{||}) = -(\tau_{z} \otimes I) D(-{\bf k}_{||}).
\end{equation}
From the condition of $b = 0$, we can finally prove~\cite{zhou2016perfect, hsu2017polarization,PhysRevLett.119.167401}
\begin{equation}
-C (\tau_{z} \otimes I)  D^{*}({\bf k}_{||}) + D({-{\bf k}_{||}}) = 0.
\label{eq_constraint1}
\end{equation}
Equation (\ref{eq_constraint1}) gives a constraint on the relative phase between the $s$ and $p$ component~\cite{hsu2017polarization}:
\begin{equation}
\text{arg} \left[ \frac{d^{\pm}_{p}({\bf k}_{||}) d^{\pm}_{p}(-{\bf k}_{||})}{d^{\pm}_{s}({\bf k}_{||}) d^{\pm}_{s}(-{\bf k}_{||})} \right]
= \text{arg} \left[ - \frac{r_{p} \pm t_{p}}{r_{s} \pm t_{s}} \right] + 2\pi N,
\label{eq_constraint2}
\end{equation}
where $d^{\pm}_{s,p} ({\bf k}_{||}) = d^{u}_{s, p} ({\bf k}_{||}) \pm d^{d}_{s,p} ({\bf k}_{||})$ and $N$ is an integer.
Equation (\ref{eq_constraint2}) give the relation between the polarization vector at ${\bf k}_{||}$ and at $-{\bf k}_{||}$.
It should be emphasized that ${\bf d}(-{\bf k}_{||})$ is \textit{not} the complex conjugate of ${\bf d}({\bf k}_{||})$.
Below the light line, an eigenmode at $-{\bf k}_{||}$ is the complex conjugate of an eigenmode at ${\bf k}_{||}$ up to an arbitrary phase factor.
Above the light line, however, non-hermiticity of the eigenvalue problem resulted from a complex frequency and an outgoing boundary condition violates this relation.
The complex conjugate of the resonance at ${\bf k}_{||}$ corresponds to a resonance at $-{\bf k}_{||}$ with a complex frequency $\omega_{0} - i\gamma$ and an incoming boundary condition.
Equation (\ref{eq_constraint2}) is a consequence of intrinsic radiation loss or the non-hermiticity of the eigenvalue problem.

The number of independent parameters in Eq.~(\ref{eq_constraint2}) is reduced by spatial symmetries.
First, consider a mirror symmetry with respect to the $xy$ plane ($\sigma_{z}$).
The mirror symmetry $\sigma_{z}$ relates the coupling constant $d^{u}_{s,p}$ with $d^{d}_{s,p}$.
The mirror operation transforms $a({\bf k}_{||}) \rightarrow a'({\bf k}_{||}) = U(\sigma_{z})a({\bf k}_{||})$ and $b({\bf k}_{||}) \rightarrow b'({\bf k}_{||}) = U(\sigma_{z})b({\bf k}_{||})$, where $U(\sigma_{z}) = \tau_{z} \otimes \mu_{x}$ and $\mu_{i}$ is the Pauli matrix.
By inserting $a'$ and $b'$ into $b'({\bf k}_{||}) = S({\bf k}_{||})a'({\bf k}_{||})$, we obtain
\begin{align}
&D({\bf k}_{||})D^{\dagger}({\bf k}_{||}) = (\tau_{z} \otimes \mu_{x}) D({\bf k}_{||})D^{\dagger}({\bf k}_{||}) (\tau_{z} \otimes \mu_{x})
\label{eq_sigma_z}
\end{align}
From Eq.~(\ref{eq_sigma_z}), we find
\begin{align}
( d^{u}_{s}({\bf k}_{||}), d^{d}_{s}({\bf k}_{||}) ) &= d_{s}({\bf k}_{||})( 1, \sigma_{z} ),
\label{eq_sigma_z_s}
\\
(d^{u}_{p}({\bf k}_{||}), d^{d}_{p}({\bf k}_{||})) &= d_{p}({\bf k}_{||})( 1, -\sigma_{z} ),
\label{eq_sigma_z_p}
\end{align}
where $\sigma_{z} = \pm 1$ is the eigenvalue of the $\sigma_{z}$ symmetry, and $\sigma_{z} = +1 (-1)$ for TE-like (TM-like) mode.
Note that the minus sign in Eq.~(\ref{eq_sigma_z_p}) results from the fact that the $\sigma_{z}$ symmetry flip the $z$-component of the electric field and does not flip the $xy$-components.
Equations (\ref{eq_sigma_z_s}) and (\ref{eq_sigma_z_p}) allow us to eliminate the superscript $u$ and $d$.
Combining Eqs. (\ref{eq_constraint1}), (\ref{eq_sigma_z_s}) and (\ref{eq_sigma_z_p}) yields~\cite{hsu2017polarization}
\begin{equation}
\text{arg}\left[ \frac{d_{p}({\bf k}_{||}) d_{p}(-{\bf k}_{||})}{d_{s}({\bf k}_{||})d_{s}(-{\bf k}_{||})} \right]
= \text{arg} \left[  \frac{-r_{p}+\sigma_{z} t_{p}}{r_{s} + \sigma_{z} t_{s}} \right] + 2 \pi N.
\label{constraint3}
\end{equation}

Next, we consider a $n$-fold rotational symmetry around the $z$ axis ($C_{n}$).
A rotational symmetry $C_{n}$ relates a resonance at $C_{n}{\bf k}_{||}$ with that at ${\bf k}_{||}$.
The $C_{n}$ operation transforms $a({\bf k}_{||}) \rightarrow a'(C_{n}{\bf k}_{||}) = a({\bf k}_{||})$, and $b({\bf k}_{||}) \rightarrow b'(C_{n}{\bf k}_{||}) = b({\bf k}_{||})$.
By inserting $a'$ and $b'$ into $b' = S(C_{n}{\bf k}_{||})a'$, we obtain
\begin{align}
D({\bf k}_{||})D^{\dagger}({\bf k}_{||}) = D(C_{n}{\bf k}_{||})D^{\dagger}(C_{n}{\bf k}_{||})
\label{eq_C_n_D}
\end{align}
From Eq.~(\ref{eq_C_n_D}), we find
\begin{align}
&D(C_{n}{\bf k}_{||}) = \frac{d^{u}_{s}(C_{n}{\bf k}_{||})}{d^{u}_{s}({\bf k}_{||})} D({\bf k}_{||}) \equiv e^{i\psi}D({\bf k}_{||}),
\label{eq_C_n}
\end{align}
where $e^{i\psi}$ is an arbitrary phase factor.
Equation (\ref{eq_C_n}) means that the polarization state at $C_{n}{\bf k}_{||}$ is same as that at ${\bf k}_{||}$.
When a PhC is invariant under $C_{2}$ operation, we can derive a constraint imposed on the polarization vector at ${\bf k}_{||}$.
By combining Eqs.~(\ref{eq_constraint1}), (\ref{eq_sigma_z_s}), (\ref{eq_sigma_z_p}) and (\ref{eq_C_n}), we derive~\cite{hsu2017polarization}
\begin{equation}
\text{arg}\biggl[ \frac{d_{p}({\bf k}_{||})}{d_{s}({\bf k}_{||})} \biggr]
= \frac{1}{2} \text{arg} \biggl[  \frac{-r_{p}+\sigma t_{p}}{r_{s} + \sigma t_{s}} \biggr] + \pi N.
\label{eq_C2T}
\end{equation}
At the $\Gamma$ point (${\bf k}_{||} = 0$), the direct scattering coefficients satisfy $r_{p} = -r_{s}$ and $t_{p} = t_{s}$, and thus the relative phase is 0 or $\pi$.
As the wavevector is away from the ${\Gamma}$ point, the relative phase deviate from $0$ or $\pi$.

Equations (\ref{eq_constraint2}), (\ref{constraint3}), and (\ref{eq_C2T}) were already derived in Ref.~\cite{hsu2017polarization}.
However, mirror operations parallel to the $z$ axis are not examined in previous works.
Next, We consider a mirror plane containing the $z$ axis ($\sigma$).
The mirror symmetry $\sigma$ relates a resonance at $\sigma {\bf k}_{||}$ with that at ${\bf k}_{||}$.
The $\sigma$ operation transforms $a({\bf k}_{||}) \rightarrow a'(\sigma {\bf k}_{||}) = - (\tau_{z} \otimes I) a({\bf k}_{||})$ and $b({\bf k}_{||}) \rightarrow b'(\sigma {\bf k}_{||}) = - (\tau_{z} \otimes I) b({\bf k}_{||})$.
By inserting $a'$ and $b'$ into $b' = S(\sigma{\bf k}_{||})a'$, we obtain
\begin{align}
    D({\bf k}_{||})D^{\dagger}({\bf k}_{||})
    = (\tau_{z} \otimes I) D(\sigma{\bf k}_{||})D^{\dagger}(\sigma{\bf k}_{||}) (\tau_{z} \otimes I)
    \label{eq_sigma_D}
\end{align}
From Eq.~(\ref{eq_sigma_D}), we find
\begin{align}
&D(\sigma {\bf k}_{||}) = \frac{d^{u}_{s}(\sigma{\bf k}_{||})}{d^{u}_{s}({\bf k}_{||})} (\tau_{z} \otimes I) D({\bf k}_{||}) \equiv e^{i\psi}(\tau_{z} \otimes I) D({\bf k}_{||}),
\label{eq_sigma}
\end{align}
which means that the relative magnitude between the $s$ and $p$ components at $\sigma {\bf k}_{||}$ is same as that at ${\bf k}_{||}$ while the relative phase at $\sigma {\bf k}_{||}$ differs from that at ${\bf k}_{||}$ by $\pi$.
At the wavevector satisfying $\sigma {\bf k}_{||} = - {\bf k}_{||}$, we can combine Eqs. (\ref{eq_constraint1}), (\ref{eq_sigma_z_s}), (\ref{eq_sigma_z_p}) and (\ref{eq_sigma}), and then find
\begin{equation}
\text{arg}\biggl[ \frac{d_{p}({\bf k}_{||})}{d_{s}({\bf k}_{||})} \biggr]
= \frac{1}{2} \text{arg} \biggl[  \frac{r_{p} - \sigma t_{p}}{r_{s} + \sigma t_{s}} \biggr] + \pi N.
\label{eq_sigmaT}
\end{equation}
Interestingly, the relative phase approaches $\pm \pi/2$ in the limit of ${\bf k}_{||} \rightarrow 0$ along $\sigma T$-invariant wavevector.
Equation (\ref{eq_sigmaT}) is the useful relation for analysis of guided resonances in PhCs without $C_{2}$ symmetry and with $\sigma T$ symmetry~\cite{liu2016pseudospin, gorlach2018far, parappurath2018direct, guo2019meron}.

\begin{figure}
\includegraphics[width=8.6cm]{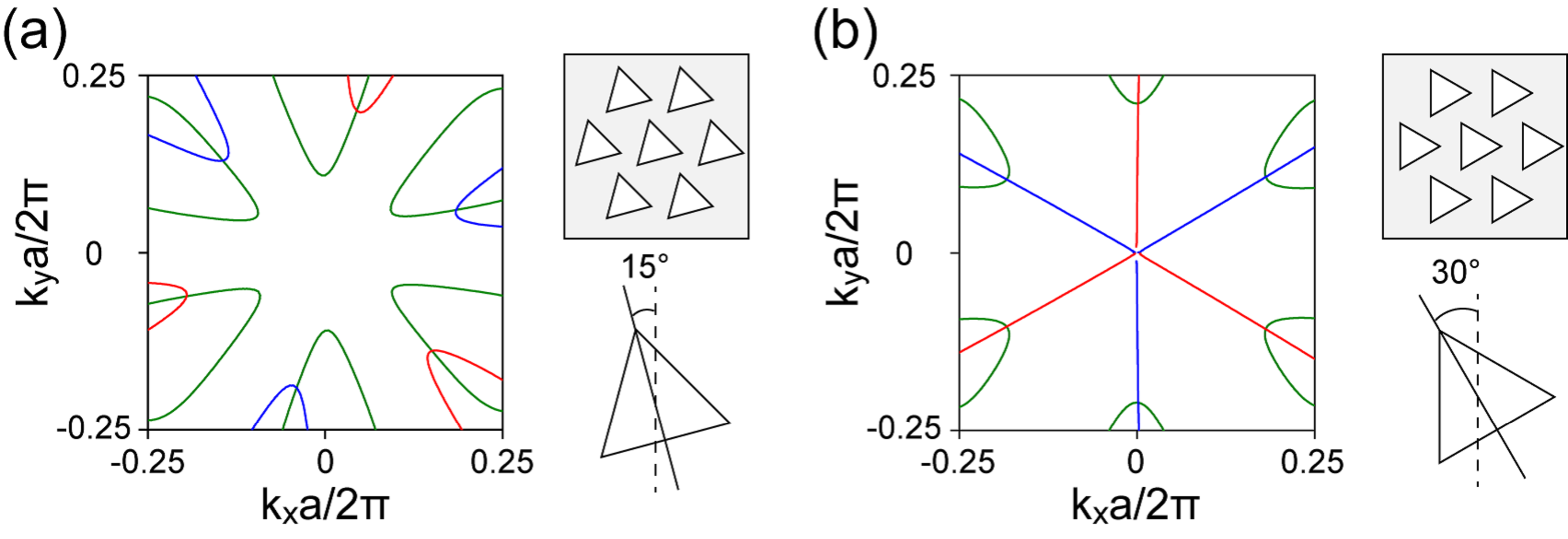}%
\caption{
(a) Lines for $|d_{s}| = |d_{p}|$ (green) and $\text{arg}(d_{p}/d_{s}) = \pm \pi/2$ (red and blue) for $\phi = 15^{\circ}$.
(b) Lines for $|d_{s}| = |d_{p}|$ and $\text{arg}(d_{p}/d_{s}) = \pm \pi/2$ for $\phi = 30^{\circ}$.
Lines for $\text{arg}(d_{p}/d_{s}) = \pm \pi/2$ is slightly deviated from $\Gamma M$ lines when $\phi = 30^{\circ}$.
The definition of line color is same in Fig.~3 in the main text.
}
\label{fig_rotated_triangle}
\end{figure}
To investigate the effect of $\sigma T$ symmetry on the polarization vector, we consider PhCs with rotated triangular holes.
When $\phi = 15^{\circ}$ [Fig.~\ref{fig_rotated_triangle}(a)], the relative phase is restricted by only Eq.~(\ref{constraint3}).
On the other hand, the relative phase is restricted by Eqs.~(\ref{constraint3}) and (\ref{eq_sigmaT}) when $\phi = 30^{\circ}$ [Fig.~\ref{fig_rotated_triangle}(b)] and $\phi = 0^{\circ}$ [Fig.~3 in the main text]. 
We can see that lines for $\text{arg}(d_{p}/d_{s}) = \pm \pi/2$ pass through the $\Gamma$ point when the in-plane mirror symmetries are present.
This is because the existence of the $\sigma T$ symmetry.
In the limit of ${\bf k}_{||} \rightarrow {\bf 0}$, the lines $\text{arg}(d_{p}/d_{s}) = \pm \pi/2$ are perpendicular to the mirror plane.

\section{S3. Analogy between polarization singularities and band degeneracies}
Although both a BIC and a CPS are singularities of $\text{arg}[S_{1} + i S_{2}]$, their winding numbers are different each other.
To show the difference of the winding of the polarization ellipse around BICs and CPSs, we examine dispersion of the Stokes parameters around these singularities.
%
%
The Stokes parameters $S_{1}$ and $S_{2}$ can be express in terms of $d'_{\pm}$ as~\cite{dennis2002polarization}
\begin{align}
\begin{split}
S_{1}({\bf k}_{||}) &= 2 \text{Re} (d'^{*}_{+} d'_{-}),
\\
S_{2}({\bf k}_{||}) &= 2 \text{Im} (d'^{*}_{+} d'_{-}).
\end{split}
\end{align}
%
When the polarization vector is circularly-polarized at ${\bf k}_{||} = {\bf k}_{0}$, the Stokes parameters $S_{1}$ and $S_{2}$ satisfy $S_{1}({\bf k}_{0}) = S_{2}({\bf k}_{0}) = 0$.
Around a CPS with $\nu_{-} = \pm 1$, we can expand the polarization vector
\begin{align}
d'_{+}({\bf k}_{0} + \delta {\bf k}_{||}) &= d'_{+}({\bf k}_{0}) + \left. \frac{\partial d'_{+}}{\partial {\bf k}_{||}} \right|_{{\bf k}_{||} = {\bf k}_{0}} \cdot \delta {\bf k}_{||},
\label{d+_expansion}
\\
d'_{-}({\bf k}_{0} + \delta {\bf k}_{||}) &= \left. \frac{\partial d'_{-}}{\partial {\bf k}_{||}} \right|_{{\bf k}_{||} = {\bf k}_{0}} \cdot \delta {\bf k}_{||}.
\label{d-_expansion}
\end{align}
By using Eqs.~(\ref{d+_expansion}) and (\ref{d-_expansion}), we can expand $S_{1}$ and $S_{2}$ around ${\bf k}_{0}$ up to the linear order of ${\bf k}_{||}$:
%
\begin{align}
S_{i}(\delta {\bf k}_{||}) = A_{ij} \delta k_{j}
\end{align}
where $\delta {\bf k}_{||} = {\bf k}_{||} - {\bf k}_{0} = (\delta k_{x}, \delta k_{y})$ is the displacement from ${\bf k}_{0}$, and $A_{ij}$ is the real constant matrix~\cite{dennis2002polarization,nye1983lines}.
The linear dependence with $\delta {\bf k}_{||}$ cause the half-winding of the polarization ellipse around ${\bf k}_{0}$; $\nu = 1/2$ for $\text{det}A > 0$ and $\nu = -1/2$ for $\text{det}A < 0$~\cite{dennis2002polarization, nye1983lines}.

Next, we assume the polarization vector vanish at ${\bf k}_{||} = {\bf k}_{0}$.
In this case, the Stokes parameters satisfy $S_{1}({\bf k}_{0}) = S_{2}({\bf k}_{0}) = 0$ and $A_{ij} = 0$ because both $d'_{+}$ and $d'_{-}$ are simultaneously zero at ${\bf k}_{0}$.
Therefore, we can expand the Stokes parameters around the BIC up to the quadratic order of ${\bf k}_{||}$:
\begin{align}
S_{i}(\delta {\bf k}_{||}) = B_{ijl}\delta k_{j} \delta k_{l},
\end{align}
%
where $B_{ijl}$ is the real constant.
In contrast to the CPS, the Stokes parameters depend on $\delta {\bf k}_{||}$ quadratically around the BIC, which causes the integer-winding of the polarization ellipse around the BIC.

The above analysis offers us an intriguing analogy between the winding of the polarization ellipse and a quantized Berry phase around band degeneracies~\cite{berry1984quantal, RevModPhys.82.1959}.
The polarization ellipse of the polarization vector is described by a real symmetric tensor~\cite{nye1983lines}
\begin{align}
	H({\bf k}_{||}) = \frac{1}{2} S_{0}({\bf k}_{||}) I - \frac{1}{2} S_{2}({\bf k}_{||}) \sigma_{x} - \frac{1}{2}S_{1}({\bf k}_{||}) \sigma_{z},
\end{align}
where $I$ is the identity matrix, and $\sigma_{i}$ is the Pauli matrices.
The two eigenvalues of $H({\bf k}_{||})$ correspond to the length of two axis of the polarization ellipses, and the two eigenvectors correspond to the direction of the two axis of the polarization ellipses when the eigenvectors are chosen to be real.
When $S_{1} = S_{2} = 0$, two eigenvalues are degenerate.
%
%
%
Because $H({\bf k}_{||})$ is real symmetric, it has same mathematical structure as a two-dimensional $PT$ symmetric Hamiltonian such as graphene~\cite{lu2014topological}.
The winding number $\nu$ corresponds to a quantized Berry phase in the two-dimensional $PT$ symmetric system, and can be written by
\begin{equation}
	2 \pi \nu = \oint_{C} d{\bf k}_{||} \cdot {\bf A}({\bf k}_{||}),
\label{eq_Berry_phase}
\end{equation}
where ${\bf A}({\bf k}_{||}) = i < u_{1} | \nabla_{{\bf k}_{||}} | u_{1} >$, and $|u_{1}>$ is the eigenvector of $H({\bf k}_{||})$ with the larger eigenvalue.
To show that Eq.~(\ref{eq_Berry_phase}) is consistent with Eq.~(5) in the main text, we perform a unitary transformation $U$ to $H({\bf k}_{||})$
\begin{align}
H'({\bf k}_{||}) &= U^{\dagger} H({\bf k}) U 
\nonumber \\
& = \frac{1}{2} S_{0}({\bf k}_{||}) I + \frac{1}{2} S_{1}({\bf k}_{||}) \sigma_{x} + \frac{1}{2}S_{2}({\bf k}_{||}) \sigma_{y},
\\
U &= \frac{1}{\sqrt{2}}
	\begin{pmatrix}
		1+i && -1-i \\
		1-i && 1-i
	\end{pmatrix}.
\end{align}
In this basis, the eigenvector of the large eigenvalue is $|u_{1}> = (1, -e^{2i\phi({\bf k}_{||})})^{T}$, where $2\phi({\bf k}_{||}) = \text{arg}[S_{1}({\bf k}_{||}) + i S_{2}({\bf k}_{||})]$, and we derive ${\bf A}({\bf k}_{||}) = \nabla_{{\bf k}_{||}} \phi({\bf k}_{||})$.
This analogy implies that the topological charge $\nu$ is finite when the loop $C$ encloses a degeneracy and that the value of the topological charge $\nu$ is determined by the dispersion of $S_{1}$ and $S_{2}$ around the degeneracy.
As mentioned above, the Stokes parameters $S_{1}$ and $S_{2}$ linearly depend on a displacement $\delta {\bf k}_{||}$ around a CPS, and quadratically depend on $\delta {\bf k}_{||}$ around a BIC, which indicates that a CPS corresponds to a linear band degeneracy and a BIC corresponds to a quadratic band degeneracy.
In this analogy, the topological charge of a CPS and a BIC corresponds to the Berry phase accumulated around the linear and quadratic band degeneracies.

\section{S4. Symmetry consideration of at-$\Gamma$ BIC}
\begin{figure}
\includegraphics[width=8.6cm]{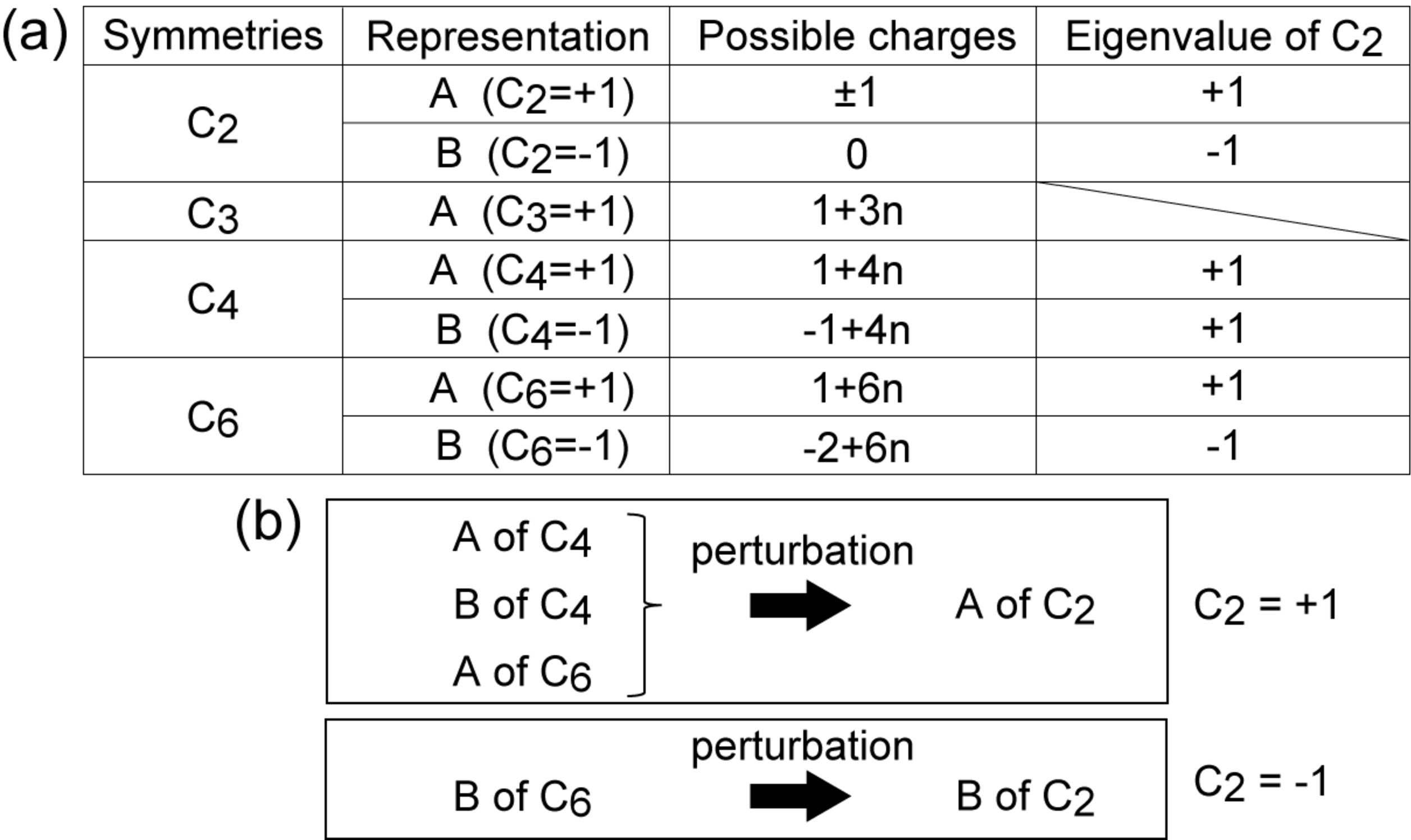}%
\caption{
(a) Possible topological charge at the $\Gamma$ point for non-degenerate bands~\cite{PhysRevLett.113.257401}.
Here $n = 0, \pm 1, \pm2, \ldots$ is an integer.
Degenerate modes with two-dimensional representations are not included.
(b) Transition of the symmetry of eigenmodes by perturbations preserving $C_{2}$ symmetry.
Eigenmodes with $C_{2} = +1 (-1)$ are reduced to the $A (B)$ representation of the $C_{2}$ point group because the perturbation preserving $C_{2}$ symmetry preserves the eigenvalue of $C_{2}$ symmetry.
}
\label{fig_symmetry}
\end{figure}

The possible charge of an at-$\Gamma$ BIC is determined by the eigenvalue of the rotational symmetry of a system as summarized in Fig.~\ref{fig_symmetry}~\cite{PhysRevLett.113.257401}.
The possible charge of eigenmodes which belong to the $B$ representation of the $C_{6}$ point group is $-2+6n$ with an integer $n$, and these modes always form at-$\Gamma$ BICs with higher-order charges independent of system parameters and mode profiles.
On the other hand, eigenmodes with belong to all other representations are allowed to be at-$\Gamma$ BICs with charge $\pm 1$.
An integer $n$ depends on the mode profile, and cannot be determined by crystalline symmetry only.
Generally, at-$\Gamma$ BICs with higher-order charge occurs at higher-frequency bands~\cite{iwahashi2011higher}.
In contrast, eigenmodes whose $C_{6}$ eigenvalue is $-1$ always form at-$\Gamma$ BICs with higher-order charges even if they are the lowest band.
For the above reason, the eigenmode with the $B$ representation of the $C_{6}$ point group is the most suitable candidate for realization of the splitting from an at-$\Gamma$ BIC into off-$\Gamma$ BICs.

\section{S5. Mobility of off-$\Gamma$ BIC}
\begin{figure}
\includegraphics[width=8.6cm]{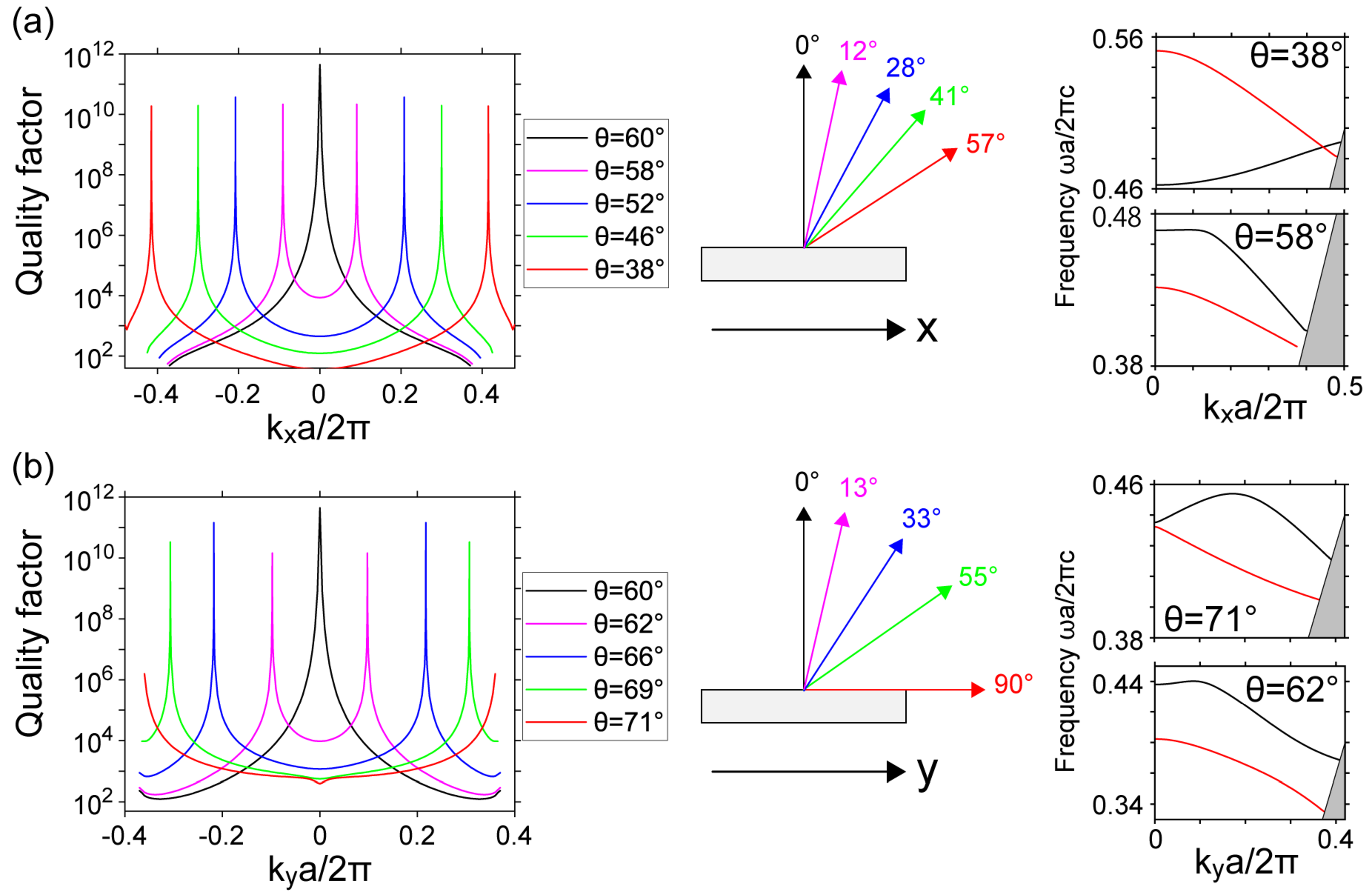}
\caption{
(a)(left) Position of the two off-$\Gamma$ BICs as a function of $\theta$.
When $\theta < 60^{\circ}$, the two BICs lies on the $k_{x}$ axis.
(middle) Outgoing angle corresponding to the BICs.
(right) The lowest two TE-like band along the $k_{x}$ axis.
We calculated the quality factor of bands marked with a red line.
When $\theta = 60^{\circ}, 58^{\circ}$, and $52^{\circ}$, the quality factor of the lowest band (red line) was calculated.
Band inversion occurs at the $\Gamma$ point while varying $\theta$.
When $\theta = 48^{\circ}$ and $38^{\circ}$, the the quality factor marked with a red line was calculated.
These two band does not repel each other on the $k_{x}$ axis because they have different mirror eigenvalues.
(b)(left) Position of the two off-$\Gamma$ BICs as a function of $\theta$.
When $\theta > 60^{\circ}$, the two BICs lies on the $k_{y}$ axis.
(middle) Outgoing angle corresponding to the BICs.
(right) The lowest two TE-like band along the $k_{y}$ axis.
We calculated the quality factor of the lowest band marked with a red line.
}
\label{fig_mobility}
\end{figure}
The off-$\Gamma$ BIC discussed in Fig.~2 can be moved by tuning system parameters.
We plot $\theta$ dependence of the two off-$\Gamma$ BIC in Fig.~\ref{fig_mobility}.
When $\theta < 60^{\circ}$, the two BICs lies on the $k_{x}$ axis.
Decreasing $\theta$ shifts the two BIC toward large wavevectors along the $k_{x}$ axis.
When $\theta > 60^{\circ}$, the two BICs lies on the $k_{y}$ axis.
Increasing $\theta$ shift the two BICs toward large wavevectors along the $k_{y}$ axis.
Eventually, the two off-$\Gamma$ BICs go outside the light cone at $\theta \approx 71^{\circ}$.
Therefore, the direction in which a radiative plane wave disappears can be varied from the vertical direction to the horizontal direction by varying $\theta$ [FIg.~\ref{fig_mobility}(b)]. 
We further emphasize that the behavior of the two off-$\Gamma$ BICs is much simpler.
Generally, it is hard to predict which direction off-$\Gamma$ BICs move in the ${\bf k}_{||}$ space when system parameters such as a slab thickness and a hole radius are varied.
In our case, however, the $C_{6}$ symmetry guarantees that the BIC must lie at the $\Gamma$ point when the perturbation is not applied.
Figure \ref{fig_mobility} imply that the BICs are away from the $\Gamma$ point as the structure of the PhC is deviated from the unperturbed structure.

The off-$\Gamma$ in Fig.~2 is fixed on the $k_{x}$ and $k_{y}$ axis by the mirror symmetries while varying $\theta$.
To move the two off-$\Gamma$ BICs two-dimensionally in the ${\bf k}_{||}$ space, we must break the mirror symmetry of the system maintaining the $C_{2}$ symmetry (e.g., changing the relative length of the two translational vectors).
Breaking the mirror symmetry can shift the two off-$\Gamma$ BIC from the mirror-invariant wavevectors~\cite{hsu2013observation}.
The direction in which a radiative plane wave disappears can be varied two-dimensionally in the ${\bf k}_{||}$ space by controlling the symmetry of a system and tuning system parameters.
These feature is suitable for directional lasing~\cite{kurosaka2010chip}.

\section{S6. Off-$\Gamma$ BIC by fine-tuning of system parameter}
\begin{figure}
\includegraphics[width=8.6cm]{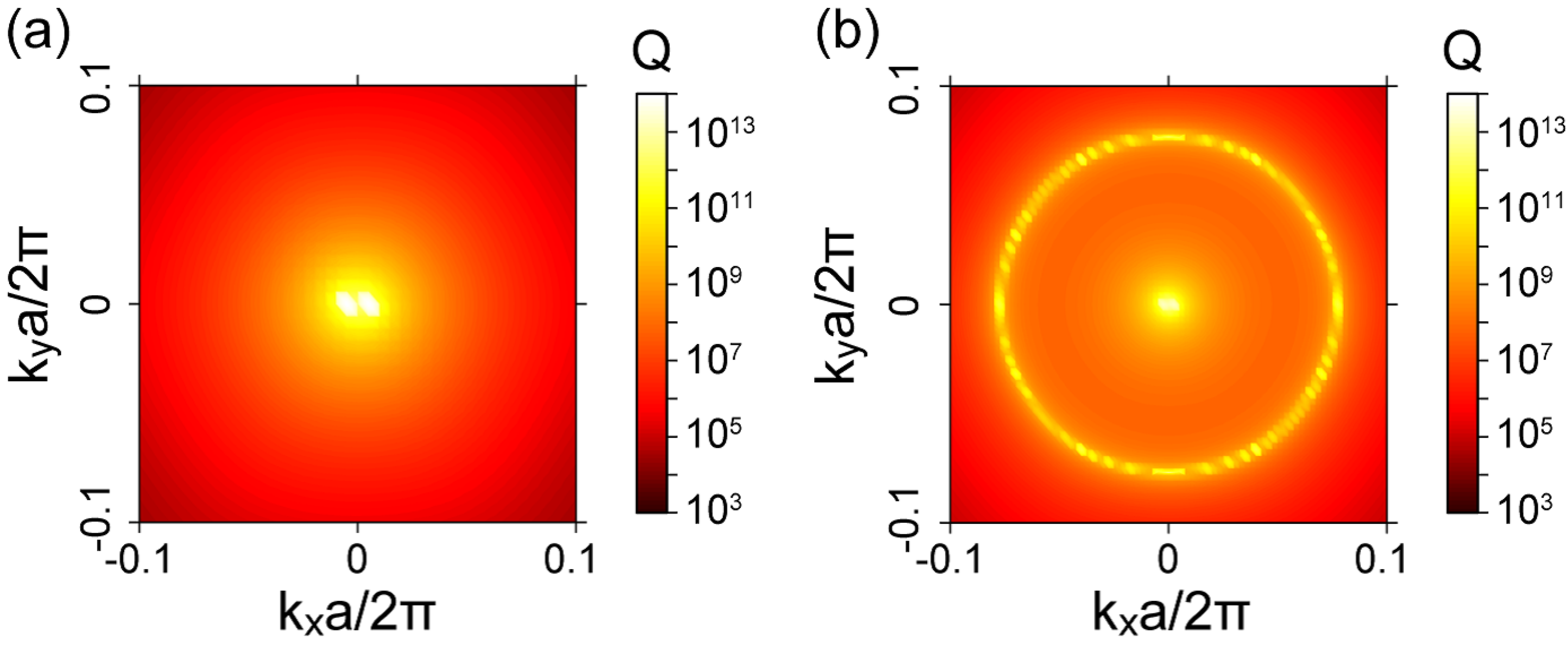}
\caption{
Calculated quality factors for (a) $h = 0.98a$ and (b) $h=a$.
All other parameters are same as that in Fig.~1 in the main text.
}
\label{fig_fine-tuning}
\end{figure}
To highlight the advantage of our work, we attempt to generate off-$\Gamma$ BICs by fine-tuning of the system parameters without symmetry breaking.
We seek off-$\Gamma$ BICs along high symmetry lines by varying the slab thickness $h$ (all other parameters are fixed).
Figure~\ref{fig_fine-tuning} shows the quality factors for $h = 0.98a$ and $h = a$.
There is no off-$\Gamma$ BICs for $h < 0.98a$ [Fig.~\ref{fig_fine-tuning}(a)].
When the slab thickness is increased, twelve off-$\Gamma$ BICs are simultaneously generated at $h \gtrsim 0.98a$, and the off-$\Gamma$ BICs lies on the $k_{x}$ axis and $k_{y}$ axis and their symmetry-related wavevectors [Fig.~\ref{fig_fine-tuning}(b)].
In contrast, the at-$\Gamma$ BIC with charge -2 always exist at the $\Gamma$ point for any parameters as long as the $C_{6}$ symmetry remains.
This demonstrates robustness of the deterministic off-$\Gamma$ BICs.
We can generate two off-$\Gamma$ BICs through our method even for $h < 0.98$ although the at-$\Gamma$ BIC is destroyed by the perturbation.